\documentclass[preprint,showpacs,preprintnumbers,amsmath,amssymb]{revtex4}
\usepackage{graphicx}
\usepackage{dcolumn}
\usepackage{bm}

\begin{document}

\title{Finite Temperature Theory of Metastable Anharmonic Potentials}

\author{ Marco Zoli }
\affiliation{Istituto Nazionale Fisica della Materia -
Dipartimento di Fisica
\\ Universit\'a di Camerino, 62032, Italy. - marco.zoli@unicam.it}

\date{\today}

\begin{abstract}
The decay rate for a particle in a metastable cubic potential is investigated in
the quantum regime by the Euclidean path integral method in
semiclassical approximation. The imaginary time formalism allows
one to monitor the system as a function of temperature. The family
of classical paths, saddle points for the action, is derived in
terms of Jacobian elliptic functions whose periodicity sets the
energy-temperature correspondence. The period of the classical
oscillations varies monotonically with the energy up to the
sphaleron, pointing to a smooth crossover from the quantum to the
activated regime. The softening of the quantum fluctuation
spectrum is evaluated analytically by the theory of the functional
determinants and computed at low $T$ up to the crossover. In
particular, the negative eigenvalue, causing an imaginary
contribution to the partition function, is studied in detail by
solving the Lam\`{e} equation which governs the fluctuation
spectrum. For a heavvy particle mass, the decay rate shows a remarkable temperature dependence
mainly ascribable to a low lying soft mode and, approaching the
crossover, it increases by a factor five over the predictions of
the zero temperature theory. Just beyond the peak value, the
classical Arrhenius behavior takes over. A similar trend is found
studying the quartic metastable potential but the lifetime of the
latter is longer by a factor ten than in a cubic potential with
same parameters. Some formal analogies with noise-induced
transitions in classically activated metastable systems are
discussed.
\end{abstract}

\pacs{03.65.Sq, 03.75.Lm, 05.30.-d,  31.15.xk}

\maketitle

\section*{1. Introduction}

Occuring in a broad class of natural phenomena metastability has
been intensively studied in different research fields ranging from
condensed matter to chemical physics, from particle and nuclear
physics to cosmology. While the mathematics of the metastable
phase in statistical physics was elaborated by Langer
\cite{langer} forty years ago, the thorough extension of the
formalism to the quantum field theory came in the seventies of the
last century \cite{volo,bender,cole}. As the semiclassical
approximation is known to be appropriate \cite{landau} to deal
with quantum tunneling, the path integral method in the Euclidean
version \cite{feyn,schul} has emerged as the fundamental tool for
quantitative theories of the decay rate $\Gamma$ of a metastable
state. The latter is given in semiclassical theory by

\begin{eqnarray}
\Gamma=\,A\exp(-B/\hbar)[1 + O(\hbar)] \label{eq:100}
\end{eqnarray}

where $A$ and $B$ depend on the specific shape of the potential.
In the path integral approach, $B$ is the Euclidean action
calculated at the classical path around which the action varies
the least while $A$ is obtained by summing over the quantum
fluctuations paths. In quadratic approximation the quantum
fluctuations are decoupled and their contribution can be evaluated
by the theory of the functional determinants which implements the
path integral method \cite{gelfand}.

Eq.~(\ref{eq:100}) is formally identical to the tunneling energy
in the bistable $\phi^4$ potential \cite{zoli} which admits an instanton
\cite{rajara,parisi} as the classical path interpolating
between the two vacua. The fundamental difference lies however in
the fact that the instanton minimizes the Euclidean action whereas
the classical path in the metastable potential, the bounce, is a
saddle point for the action. Hence the bounce time derivative,
which is eigenfunction of the quantum fluctuations Sturm-Liouville
operator with zero eigenvalue (due to time translational
invariance), is not the ground state. There must be a negative
eigenvalue \cite{affl} in the fluctuation spectrum causing
an imaginary contribution to the total partition function. The
decay rate of the false vacuum arises precisely from this negative
eigenvalue which, moreover, is unique as the bounce
path has one node along the time axis.

While the general formalism of metastability is settled at least
in one dimension and in the semiclassical method \cite{weisshaef,volo1}, the
evaluations of $\Gamma$ for specific systems are generally based
on the Langer-Coleman approach \cite{kleinert} which assumes a
system with {\it infinite size} $L$  and therefore, strictly
speaking, holds in the zero temperature limit of the quantum field
theory once $L$ is taken equal to $\hbar c/K_BT$. The same formal
approach can be used in classical systems to calculate the
noise-induced transition rate between locally stable states, for
example the domain reversal rate in micromagnets \cite{braun} and
the instabilities in metallic nanowires \cite{stein1}. In such
systems however the character of the transition between activation
regimes crucially depends on the finite spatial extent
\cite{faris}, thus quantitative estimates of the Kramer escape
rate require an extension of the Langer-Coleman method to the more
complex {\it finite size} case. In quantum tunneling systems, the
counterpart of the latter is represented by the {\it finite T}
quantum regime below the crossover temperature $T^*_{c}$ at which
thermal activated processes set in. Below $T^*_{c}$, the decay
rate stems from purely quantum fluctuations effects and the
$\Gamma$ dependence on $T$ is governed by the softening of the
fluctuation spectrum. To determine this phenomenon with accuracy
one has to face the non trivial task of computing fluctuation
functional determinants at low but finite $T$ \cite{n1} which also
includes a knowledge of the negative eigenvalue in the spectrum.
This paper focuses on this issues developing the finite temperature
theory of metastability below $T^*_{c}$ for the  case of a model
cubic potential.  In particular, the softening of the fluctuations is analysed
quantitatively by means of the Jacobi elliptic functions formalism
\cite{wang} which has the advantage to monitor the evolution of
the system at any $T$, free from approximations regarding the form
of the bounce and its fluctuation spectrum. The explicit equations
which determine such softening are provided. Activation effects
\cite{hanggi,antunes} are not discussed in this work which assumes a
closed system. Metastability in the presence of dissipation has
been studied in Ref.\cite{rise}.

As in the space-time Euclidean path integral the time is an
inverse temperature, the zero $T$ theory assumes that the bounce
classical motion lasts an {\it infinite time} whereas the finite
$T$ theory realistically admits that the excursion time for the
bounce may be finite. This consideration permits to derive the
family of energy dependent paths which solve the classical
equation of motion as shown in Section 2 for the cubic potential.
In Section 3, using the theory of the functional determinants, I
determine analytically and compute the quantum fluctuations
contribution to the path integral. The solution
of the Lam\`{e} equation which governs the fluctuation spectrum is presented
together with the evaluation of the temperature effects on the lowest lying
eigenmodes and their eigenvalues. Section 4 contains the results
for the decay rate of the metastable cubic potential and a
comparison with the case of the metastable quartic potential. Some
final remarks are made in Section 5.

\section*{2. Cubic Potential Model}

Consider a particle of mass $M$ in the one dimensional cubic
potential:

\begin{eqnarray}
V(x)=\, {{M\omega^2} \over 2}x^2 - {{\gamma} \over 3}x^3
\label{eq:55}
\end{eqnarray}

plotted in Fig.~\ref{fig:1}(a) for several energies $\hbar\omega$.
Say $a$ the position of the top of barrier whose height is
$V(a)=\,\gamma a^3/6$ with $\gamma=\,M\omega^2/a$. I take
throughout the paper, $a=\,1\AA$ and $M=\,10^4m_e$ with $m_e$
being the electron mass. At $x=\,0$ the particle is in a local
minimum, see Fig.~\ref{fig:1}(a), from which it cannot escape
classically and, in the real time formalism, only $x_{cl}=\,0$
solves the classical equation of motion. Such local minimum is
however metastable as quantum fluctuations allow the particle to
explore the abyss at $x \geq 3a/2$. Let's see how.

\subsection*{A. Classical Path}

\begin{figure}
\includegraphics[height=7cm,angle=-90]{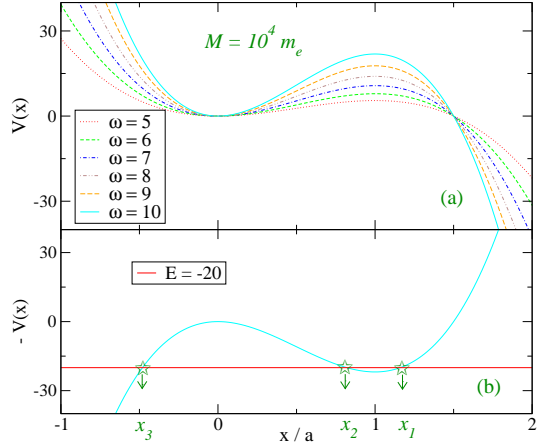}
\caption{\label{fig:1}(Color online) (a) Cubic potential (in
$meV$) for several oscillator energies, (b) Cubic potential for
$\hbar\omega=\,10meV$ in the imaginary time representation. The
intersections with the constant energy $E$ (in $meV$) define the
turning points for the classical motion.}
\end{figure}

A non trivial classical solution can be found in the Euclidean
space. In fact, performing a Wick rotation from the real to the
imaginary time, $t \rightarrow -i\tau$, the equation of motion
reads:

\begin{eqnarray}
M\ddot{x}_{cl}(\tau)=\,V'(x_{cl}(\tau)) \label{eq:1}
\end{eqnarray}

where $V'$ means derivative with respect to $x_{cl}$. The particle
path $x(\tau)$ has been split in the sum of a classical and a
quantum component, $x(\tau)=\,x_{cl}(\tau) + \eta(\tau)$. The Wick
rotation is equivalent to turn the potential upside down with
respect to the real time as shown in Fig.~\ref{fig:1}(b) for
$\hbar\omega=\,10meV$. In the reversed potential the classical
motion can take place.

Integrating Eq.~(\ref{eq:1}), one gets:

\begin{eqnarray}
{M \over 2}\dot{x}_{cl}^2(\tau) - V(x_{cl}(\tau))=\, E
\label{eq:2}
\end{eqnarray}

with the constant $E$ representing the classical energy. Defining:

\begin{eqnarray}
& &\chi_{cl}(\tau)=\,
{2 \over {3}}{{x_{cl}(\tau)} \over a} \,
\nonumber
\\
& &\kappa=\,{{4E}\over {27 V(a)}} \label{eq:56}
\end{eqnarray}

Eq.~(\ref{eq:2}) is easily integrated to yield:

\begin{eqnarray}
\tau - \tau_0 =\,\pm {{1 \over \omega}}
\int_{\chi_{cl}(\tau_0)}^{\chi_{cl}(\tau)} {{d\chi} \over
{\sqrt{-\chi^3 + \chi^2 + \kappa }}} \label{eq:57}
\end{eqnarray}

where $\tau_0$ is the center of motion  between the turning points
at which the path velocity vanishes. Thus the physical picture set
by the boundary conditions is that, see Fig.~\ref{fig:1}(b), of a
particle starting from $x_2$ at the time $\tau=\,-L/2$, reaching
$x_1$ at $\tau=\,\tau_0$ and returning to the initial position at
$\tau=\,L/2$. Then, Eq.~(\ref{eq:57}) has a time reversal
invariant solution whose period $L$ is finite and dependent on
$E$. Phase transitions phenomena in spatially extended systems are
analysed by formally similar models with $L$ being the size of the
system \cite{faris,stein2}. Looking at Fig.~\ref{fig:1}(b), one
captures qualitatively that the amplitude $x_1 - x_2$ attains the
largest value for the $E=\,0$ motion while $x_2$ and $x_3$
coincide. $x_3$ becomes negative for $E < \,0$ motions.

The turning points $x_1$, $x_2$, $x_3$ are given by the zeros of
the equation $-\chi^3 + \chi^2 + \kappa=\,0$ ($\chi\equiv \,
2x/(3a)$) which admits three real solutions for $\kappa \in
[-4/27, 0]$, that is for $E \in [-V(a),0]$. After some algebra I
find:

\begin{eqnarray}
& &\chi_1=\, {1 \over {3}} + {2 \over {3}}\cos(\vartheta) \,
\nonumber
\\
& & \chi_2=\, {1 \over {3}} + {2 \over {3}}\cos(\vartheta -
2\pi/3) \, \nonumber
\\
& & \chi_3=\, {1 \over {3}} + {2 \over {3}}\cos(\vartheta -
4\pi/3) \, \nonumber
\\
& & \vartheta=\,{1 \over {3}}\arccos\bigl({{27 \kappa} \over {2}}
+ {1}\bigr) \label{eq:58}
\end{eqnarray}

At the bounds of the energy range, Eq.~(\ref{eq:58}) yields:

\begin{eqnarray}
& &{\bf  E=\,0}  \Rightarrow  \chi_1=\,1 ;\,{}\, \chi_2=\,
\chi_3=\,0 \, \nonumber
\\
& & {\bf  E=\,-V(a)} \Rightarrow  \chi_1=\,\chi_2=\,2/3 ;\,{}\,
\chi_3=\,-1/3 \, \nonumber
\\
\label{eq:59}
\end{eqnarray}

Thus, at the sphaleron energy $E_{sph}=\,|E|=\,V(a)$
\cite{manton}, the amplitude of the finite time solution has to
shrink into a point. Let's find the general solution of
Eq.~(\ref{eq:57}) by pinning the center of motion at
$\chi_{cl}(\tau_0)=\,\chi_1$ and using the result \cite{grad}:

\begin{eqnarray}
& &\int_{\chi_{cl}(\tau)}^{\chi_{1}} {{d\chi} \over {\sqrt{(\chi_1
- \chi)(\chi - \chi_2)(\chi - \chi_3) }}}=\,{{2 F(\lambda,p)}
\over {\sqrt{\chi_1 - \chi_3}}} \, \nonumber
\\
& &\lambda=\,\arcsin\Biggl(\sqrt{{\chi_1 - \chi_{cl}(\tau)}\over
{\chi_1 - \chi_2}}\Biggr)\, \nonumber
\\
& &p=\, \sqrt{{\chi_1 - \chi_{2}}\over {\chi_1 - \chi_3}}
\label{eq:61}
\end{eqnarray}

where $F(\lambda,p)$ is the elliptic integral of the first kind
with amplitude $\lambda$ and modulus $p$.  Then, through
Eqs.~(\ref{eq:56}),~(\ref{eq:57}),~(\ref{eq:61}), I derive the
bounce solution of the {\it finite time} theory:

\begin{eqnarray}
& &x_{cl}(\tau)=\,{{3a} \over 2}\bigl[\chi_1 cn^2(\varpi,p) +
\chi_2 sn^2(\varpi,p)\bigr] \, \nonumber
\\
& &\varpi=\, \sqrt{{\chi_1 - \chi_3}}\,{\omega \over 2}(\tau -
\tau_0)\, \nonumber
\\
\label{eq:62}
\end{eqnarray}

$sn(\varpi,p)$ and $cn(\varpi,p)$ are the {\it sine-} and {\it
cosine-} amplitudes respectively \cite{wang}. The modulus $p$
incorporates the classical mechanics of the problem through the
second of Eq.~(\ref{eq:56}) and Eq.~(\ref{eq:58}). At {E =\,0},
$p=\,1$, the bounce of the {\it infinite time} theory is
recovered:

\begin{eqnarray}
& &x_{cl}(\tau)=\,{{3a} \over 2}cn^2(\varpi,1)=\,{{3a} \over 2}
sech^2\bigl({\omega \over 2}(\tau - \tau_0)\bigr) \, \nonumber
\\
\label{eq:63}
\end{eqnarray}

\begin{figure}
\includegraphics[height=7cm,angle=-90]{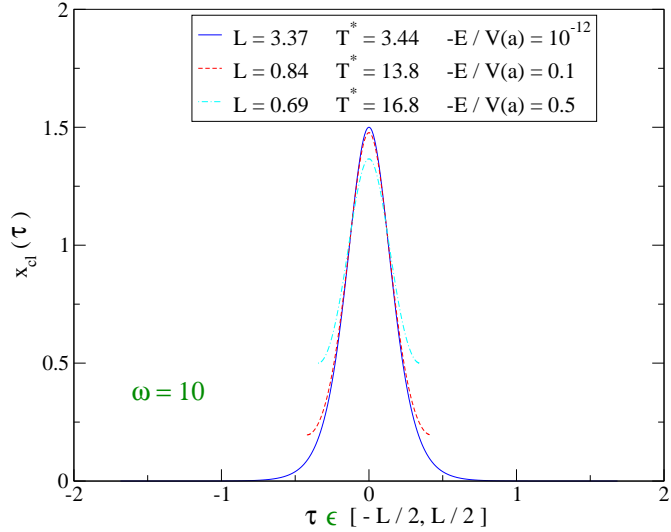}
\caption{\label{fig:2}(Color online) Shape of the bounce solution
for three classical energies $E$. The oscillation period $L/\hbar$
is in units $meV^{-1}$ and $T^*$ in $K$.}
\end{figure}

At the sphaleron energy, $p=\,0$,  the bounce solution is in fact
a point-like object set at the bottom of the valley in the
reversed potential: $x_{cl}(\tau)=\,a$. Thus, Eq.~(\ref{eq:62})
defines the transition state which is a saddle for the action
below the sphaleron. Representing the Jacobi elliptic functions in
Eq.~(\ref{eq:62}) through trigonometric series \cite{grad}
suitable to computation, I plot the classical path in
Fig.~\ref{fig:2} which makes evident how the bounce amplitude
contracts by increasing the {\it energy over potential height}
ratio (in absolute value). While the bounce can be physically
interpreted as a pair of domain walls, it is clear that the extent
of the $\tau$-region over which the domain walls separate the
turning points continuously shrinks by approaching the crossover.
The center of motion $\tau_0$ has been set equal to zero with no
loss of generality being the system invariant for time
translation.

As the bounce is a combination of squared Jacobi elliptic
functions, its period is $2K(p)$ with $K(p)=\,F(\pi/2,p)$ being
the complete elliptic integral of the first kind \cite{wang}.
Hence, from Eq.~(\ref{eq:62}), I get:

\begin{eqnarray}
\sqrt{{\chi_1 - \chi_3}}\,{\omega \over 4}L=\,K(p) \label{eq:64}
\end{eqnarray}

which establishes the relation between the oscillation period and
the classical energy embedded in the turning points. As in the
Matsubara formalism, one can map the imaginary time onto the
temperature axis, $L=\,\hbar /(K_BT^*)$, where $T^*$ is the
temperature at which the particle makes the journey to and from
the edge of the abyss for a given $E$.  Then, only periodic
bounces whose period equals the inverse temperature determine the
decay rate and the {\it finite time} theory can be viewed as a
finite $T^*$ theory. From Eq.~(\ref{eq:64}) I get:

\begin{eqnarray}
K_BT^*=\,{{\hbar \omega} \over 4} {\sqrt{{\chi_1 - \chi_3}} \over
{ K(p)}}\label{eq:64a}
\end{eqnarray}

The $L$ and $T^*$ values given in Fig.~\ref{fig:2} are computed on
the base of Eqs.~(\ref{eq:64}, ~(\ref{eq:64a}) respectively.

At the sphaleron, Eq.~(\ref{eq:64a}) yields

\begin{eqnarray}
K_BT_{c}^*=\,{{\hbar \omega} \over {2\pi}} \label{eq:64b}
\end{eqnarray}

that marks the transition temperature between quantum and
activated regimes. Such value represents the upper bound for the
validity of the model and precisely sets the Goldanskii criterion
\cite{gold} for the cubic potential. Taking $\hbar
\omega=\,10meV$, the following calculations are carried out in the
low temperature range up to $T_{c}^*=\,18.469K$.

\subsection*{B. Classical Action}

The classical action $A[x_{cl}]$ for the bounce in the finite temperature
theory can be computed either in terms of the path velocity
$\dot{x}_{cl}(\tau)$:

\begin{eqnarray}
& &A[x_{cl}]=\, 2M \int_{-L/2}^{\tau_0}d\tau
[\dot{x}_{cl}(\tau)]^2 - E\cdot L(E) \, \nonumber
\\
& & \dot{x}_{cl}(\tau)=\,{{3a} \over 2}\mathcal{F}\cdot
sn(\varpi,p)cn(\varpi,p)dn(\varpi,p)\, \nonumber
\\
& &\mathcal{F}=\,-\omega (\chi_1 - \chi_2) \sqrt{{\chi_1 -
\chi_3}} \label{eq:65}
\end{eqnarray}

($dn(\varpi,p)$ is the {\it delta-} amplitude \cite{wang}) or, in
terms of the potential $V(x_{cl})$:

\begin{eqnarray}
& &{{A[x_{cl}]}}=\, {{27V(a)} \over {\omega}}
\int_{\chi_2}^{\chi_1}d\chi \sqrt{{\kappa} + \chi^2 - \chi^3} -
{E\cdot L(E)} \, \nonumber
\\
\label{eq:65a}
\end{eqnarray}

Both ways require computation of $L(E)$ through
Eqs.~(\ref{eq:56}),~(\ref{eq:58}),~(\ref{eq:64}).

\begin{figure}
\includegraphics[height=7cm,angle=-90]{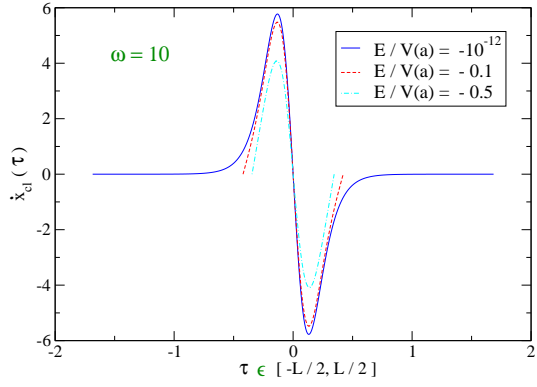}
\caption{\label{fig:3}(Color online) Bounce velocity versus
imaginary time for the same parameters as in Fig.~\ref{fig:2}. }
\end{figure}

\begin{figure}
\includegraphics[height=7cm,angle=-90]{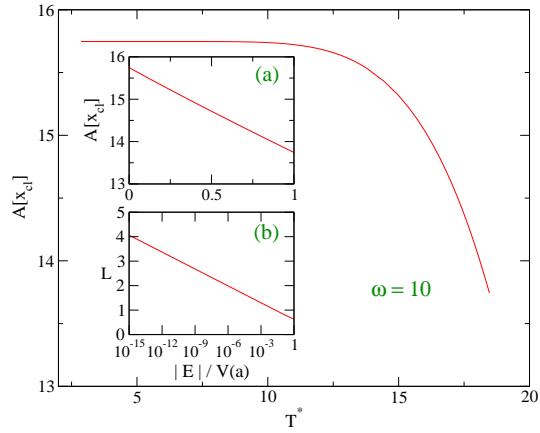}
\caption{\label{fig:4}(Color online) Classical action (in units
$\hbar$) versus temperature. Insets: (a) Classical Action versus
{\it Energy over Barrier potential height} ratio on a linear
scale; (b) Oscillation period (over $\hbar$) versus {\it Energy
over Barrier potential height} ratio on a log scale.}
\end{figure}

In the $E\rightarrow 0$ limit, I get the result

\begin{eqnarray}
{{A[x_{cl}]} \over \hbar} \rightarrow \,  {{6 M^3 \omega^5} \over
{5\hbar \gamma^2}} \label{eq:66}
\end{eqnarray}

which serves as testbench for the computational method. The
inverse dependence on the anharmonic force constant $\gamma$
reflects the well known fact that metastable systems are
intrinsecally non perturbative and provides the fundamental
motivation for the semiclassical treatment. Eq.~(\ref{eq:66})
permits to set the potential parameters such as the condition
${A[x_{cl}] \gg \hbar}$ holds and the semiclassical method is thus
justified. As $M$ and $a$ have been taken constant, ${{A[x_{cl}]}
\over \hbar}\propto \omega$ in the $E\rightarrow 0$ limit. The
bounce velocity and the classical action are displayed in
Fig.~\ref{fig:3} and Fig.~\ref{fig:4} respectively. Note that
${A[x_{cl}]}$ decreases smoothly with increasing temperature and
absolute value of energy (inset (a) in Fig.~\ref{fig:4})
suggesting that the transition to the activated regime above
$T_{c}^*$ is of second order \cite{affl,larkin}. Although I'm
dealing here with a closed system, it is worth emphasizing that
the inclusion of a dissipative environment \cite{grabwei} would
decrease $T_{c}^*$ but would not change the character of the
crossover.

In general, both first and second order transitions have been
found in tunneling systems depending on the shape of the potential
\cite{blatter,kuznet,liang}. The criterion used to establish the
order of the transitions in periodic tunneling systems has been
formulated by Chudnovsky \cite{chudno} on the base of the behavior
of $L(E)$: a monotonic dependence of the period below the
sphaleron points to a smooth crossover whereas a nonmonotonic
$L(E)$ indicates a sharp transition. The inset (b) in
Fig.~\ref{fig:4} clearly shows that the former case applies to the
cubic potential in Eq.~(\ref{eq:55}). $L(E)$ decreases
monotonically and, consistently, the action versus $T^*$ is convex
upwards. At the sphaleron, I find numerically:
$L(E_{sph})/\hbar=\,0.628meV^{-1}$. Note that such value
corresponds to $2\pi/\hbar\omega$ \cite{n3} and this is not by
accident. In fact at the sphaleron the bounce is a point, that is
a static solution of Eq.~(\ref{eq:2}) but, near the sphaleron, the
periodic path is the sum of the sphaleron and an oscillation with
negative eigenvalue $\varepsilon_{-1}$ whose period tends to
$L(E_{sph})=\,2\pi/\sqrt{|\varepsilon_{-1}|}$ \cite{park}. Then
one infers that, for $|E| \rightarrow E_{sph}$,
$\varepsilon_{-1}\rightarrow -\omega^2$. I'll come back to this
point towards the end of the next Section.

\section*{3. Quantum Fluctuations}

Let's expand the action around the classical path up to second
order in the fluctuations

\begin{eqnarray}
& &A[x]\sim \,A[x_{cl}] + {\delta} A[\eta] + {1 \over {2!}}
{\delta}^2 A[\eta]\, \nonumber
\\
& & {\delta} A[\eta]=\, \int_{-L/2}^{L/2} d\tau \biggl({M}
\dot{x}_{cl}(\tau) \dot{\eta}(\tau) + V'(x_{cl}(\tau)){\eta}(\tau)
\biggr) \, \nonumber
\\
& &{\delta}^2 A[\eta]\equiv \,2 A_f[\eta] \, \nonumber
\\
& &A_f[\eta]=\,\int_{-L/2}^{L/2} d\tau \biggl({M \over 2}
\dot{\eta}^2(\tau) + {1 \over 2}V''(x_{cl}(\tau))\eta^2(\tau)
\biggr) \, \nonumber
\\
\label{eq:66+}
\end{eqnarray}

Being the classical solution an extremum for the action, the first
functional derivative vanishes and, after partial integration in
the first term of ${\delta} A(\eta)$, one gets the Euler-Lagrange
Eq.~(\ref{eq:1}).

Further, differentiating Eq.~(\ref{eq:1}) with respect to $\tau$,
one observes that ${\dot{x}_{cl}(\tau)}$ solves the homogeneous
equation associated to the second order Schr\"{o}dinger-like
differential operator in $A_f[\eta]$

\begin{eqnarray}
& &\hat{O} \eta_n(\tau)=\,\varepsilon_n \eta_n(\tau) \, \nonumber
\\
& &\hat{O}\equiv -\partial_{\tau}^2 + {{ V''(x_{cl}(\tau))}/
M}\,\nonumber
\\
& &\eta(\tau)=\,\sum_{n=\,-1}^{\infty} \varsigma_n \eta_n(\tau)
\label{eq:66++}
\end{eqnarray}

where the $\varepsilon_n$ are the fluctuations eigenvalues while
the coefficients $\varsigma_n$ of the series expansion in
ortonormal components $\eta_n(\tau)$ define the measure of the
fluctuation paths integration.  Then, as a general consequence of
the $\tau$-translational invariance, ${\dot{x}_{cl}(\tau)}$ is
proportional to the eigenmode $\eta_0(\tau)$ with
$\varepsilon_0=\,0$.  The latter cannot be the ground state as,
see Fig.~\ref{fig:3}, ${\dot{x}_{cl}(\tau)}$ has one node hence,
the quantum fluctuation ground state has negative eigenvalue. From
Eq.~(\ref{eq:65}), note that for any two points $\varpi_1,
\varpi_2$ such that $\varpi_2=\,\varpi_1 \pm 2K(p)$,
$\dot{x}_{cl}(\varpi_2)=\,\dot{x}_{cl}(\varpi_1)$. The important
consequence is that the fluctuation eigenmodes obey periodic
boundary conditions (PBC).

After these observations we are ready to calculate the space-time
Euclidean path integral between the positions $x_i$ and $x_f$
connected in the time $L$. In the semiclassical model, it is given
by

\begin{eqnarray}
& &<x_f|x_i>_L=\,\exp\biggl[- {{A[x_{cl}]} \over {\hbar}} \biggr]
\cdot \int D\eta \exp\biggl[- {{A_f[\eta]} \over {\hbar}}
 \biggr] \, \nonumber
\\
& &\int D\eta=\,\aleph \prod_{n=\,-1}^{\infty}
\int_{-\infty}^{\infty} {{d\varsigma_n}\over {\sqrt{2\pi\hbar/M}}}
\label{eq:66+++}
\end{eqnarray}

$\aleph$ depends only on the functional integral measure. $x_i$
and $x_f$ coincide for the periodic bounce, thus
Eq.~(\ref{eq:66+++}) represents the single bounce contribution
$Z_1$ to the total partition function $Z_T$. The latter also
contains the effects of all multiple (non interacting) excursions
to and from the abyss which is equivalent to sum \cite{schul} over
an infinite number of single bounce contributions like $Z_1$.
Moreover, also the static solution of Eq.~(\ref{eq:1}),
$x_{cl}=\,0$ contributes with the harmonic partition function
$Z_h$ which can be easily determined using the measure in
Eq.~(\ref{eq:66+++}). Summing up, $Z_T$ is given by

\begin{eqnarray}
& & Z_T=\,Z_h \exp(Z_1/Z_h)\, \nonumber
\\
& &Z_h=\,\aleph |Det[\hat{h}]|^{-1/2} \label{eq:66a+++}
\end{eqnarray}

$Det[\hat{h}]$ being the harmonic fluctuation determinant. Being
the decay rate $\Gamma$ proportional to the imaginary exponential
argument through the Feynman-Kac formula \cite{schul}, it follows
that there is no need to determine $\aleph$ as it cancels out in
the ratio $Z_1/Z_h$.

Let's proceed to evaluate the fluctuation term in $Z_1$ which,
formally, is worked out by carrying out Gaussian path integrals
yielding:

\begin{eqnarray}
& &\int D\eta \exp\biggl[- {{A_f[\eta]} \over {\hbar}} \biggr]= \,
\aleph \cdot Det\Bigl[\hat{O}\Bigr]^{-1/2} \, \nonumber
\\
& & Det[\hat{O}]\equiv \, \prod_{n=\,-1}^{\infty}\varepsilon_n
\label{eq:66++++}
\end{eqnarray}

Eq.~(\ref{eq:66++++}) is however divergent due to the Goldstone
mode which reflects the fact that $\tau_0$, the center of the
bounce, can be located arbitrarily inside $L$. However such mode
can be treated separately \cite{larkin}: the divergent integral
over the coordinate $d\varsigma_0$ associated to the zero mode in
the measure $D\eta$ is transformed into a $d\tau_0$ integral.
Accordingly the eigenvalue $\varepsilon_0=\,0$ is extracted from
$Det[\hat{O}]$ and its contribution to Eq.~(\ref{eq:66++++}) is
replaced as follows

\begin{eqnarray}
& & (\varepsilon_0)^{-1/2} \rightarrow \sqrt{{{M}\over
{2\pi\hbar}}}\bar{N}^{-1}L\, \nonumber
\\
& & \bar{N}^{-1}=\,\sqrt{ 2\int_0^{L/2}d\tau
|\dot{x}_{cl}(\tau)|^2}
\label{eq:66+++++}
\end{eqnarray}

To be rigorous, this replacement holds in the approximation of
quadratic fluctuations \cite{kleinert} while higher order terms
may be significant around the crossover. It is also worth noticing
that Eq.~(\ref{eq:66+++++}) is often encountered in the form
$(\varepsilon_0)^{-1/2} \rightarrow \sqrt{{{A[x_{cl}]}\over
{2\pi\hbar}}}L$. However, this is correct only in the zero $T$
limit where $A[x_{cl}]$ equals $M \bar{N}^{-2}$ (see
Eq.~(\ref{eq:65})) while, approaching $T^*_c$, the difference
between the two objects gets large. Exactly at $T^*_c$, the norm
of the bounce velocity vanishes while the action is finite. This
fact is crucial in establishing the behavior of the decay rate at
the crossover as shown below.

Now I tackle the problem of the evaluation of the regularized
determinant $Det^R[\hat{O}]$ defined by
$Det[\hat{O}]=\,\varepsilon_0 \cdot Det^R[\hat{O}]$.

\subsection*{A. Fluctuation Determinant}

The calculation of $Det^R[\hat{O}]$ requires knowledge only of the
classical paths. This is a fundamental feature of the theory of
functional determinants of second order differential operators
first developed for Dirichlet boundary conditions \cite{gelfand}
and then extended to general operators and boundary conditions in
several ways \cite{forman,kirsten1}. As shown
above the path velocity obeys PBC for any two points $\varpi_1,
\varpi_2$ separated by the period $2K(p)$. The latter corresponds
to the oscillation period $L$ along the $\tau$-axis. It can be
easily checked that also the path acceleration fulfills the PBC.
Then, the regularized determinant is given by

\begin{eqnarray} Det^R[\hat{O}]=\,{{<f_0 |f_0> \bigl(f_1(\varpi_2) -
f_1(\varpi_1)\bigr)} \over {{f_0(\varpi_1) W(f_0, f_1)}}}
\label{eq:67}
\end{eqnarray}

where $f_0, f_1$ are two independent solutions of the homogeneous
equation: $\hat{O} \eta_n(\tau)=\,0$. $W(f_0, f_1)$ is their
Wronskian and $<f_0 |f_0>$ is the squared norm. $f_0$ is obviously
$\dot{x}_{cl}$ while $f_1$ can be taken as

\begin{eqnarray}
f_1=\,{{\partial {x}_{cl}} \over {\partial q}}\, ;{}\, q \equiv
\,p^2 \label{eq:67+}
\end{eqnarray}

Using Eq.~(\ref{eq:62}), derivatives and properties of the
elliptic functions, I derive for $f_1$:

\begin{eqnarray}
f_1&=&\,{{3a} \over{2}} \Biggl[{{\partial \chi_1} \over {\partial
q}} cn^2(\varpi,p) + {{\partial \chi_2} \over {\partial q}}
sn^2(\varpi,p)  \, \nonumber
\\ & &- 2(\chi_1 - \chi_2)
sn(\varpi,p)cn(\varpi,p)dn(\varpi,p) \cdot {{\partial F(\lambda,
p)} \over {\partial q}}\Biggr]\, \nonumber
\\
{{\partial \chi_1} \over {\partial q}}&=&\, {{1 - p^2} \over {2(1
- p^2 + p^4)^{3/2}}}\, \nonumber
\\
{{\partial \chi_2} \over {\partial q}}&=&\, {{-1} \over {2(1 - p^2
+ p^4)^{3/2}}}\, \nonumber
\\
{{\partial F(\lambda,p)}\over {\partial q}}&=&\,{1 \over {
2\bar{p}^2}}\Biggl[ {{E(\lambda,p) - \bar{p}^2 F(\lambda,p)} \over
{p^2}} - {{\sin\lambda \cos\lambda} \over {\sqrt{1 -
p^2\sin^2\lambda}}} \Biggr]\, \nonumber
\\
& &\bar{p}^2=\, 1 - p^2 \label{eq:68}
\end{eqnarray}

where $E(\lambda,p)$ is the elliptic integral of the second kind.

The Wronskian, being constant along $\tau$, can be calculated in
any convenient point. Let's take $\tau_0$ as $f_0(\tau_0)=\,0$.
Then:

\begin{eqnarray}
& &W(f_0, f_1)\Bigr|_{\tau_0}=\,- \dot{f}_0(\tau_0)f_1(\tau_0) \,
\nonumber
\\
& &=\,{9 \over 8}a^2\omega^2 (\chi_1 - \chi_2){(\chi_1 - \chi_3)}
{{\partial \chi_1} \over {\partial q}} \, \nonumber
\\
\label{eq:69}
\end{eqnarray}

and $<f_0 | f_0>\equiv \bar{N}^{-2}$ given in
Eq.~(\ref{eq:66+++++}).

Thus, all ingredients are known to calculate ${Det^R[\hat{O}]}$:
evaluating Eq.~(\ref{eq:68}) on the $\tau-$ range boundaries,
Eq.~(\ref{eq:67}) transforms into

\begin{eqnarray}
{Det^R[\hat{O}]} =\,{{2} \over {\omega \sqrt{\chi_1 - \chi_3}
\bar{p}^2}} \Biggl[ {{E(\pi/2,p) - \bar{p}^2K(p)} \over {p^2}}
\Biggr] \cdot {{<f_0 |f_0>} \over {W(f_0, f_1)}} \label{eq:70}
\end{eqnarray}

which can be computed using
Eqs.~(\ref{eq:66+++++}),~(\ref{eq:69}). It is however known in the
theory of functional determinants \cite{gelfand} that only ratios
of determinants are meaningful in value and sign, such ratios
arising naturally in the path integral method as it has been
pointed out above. In fact, $Det^R[\hat{O}]$ would diverge in the
$T^* \rightarrow 0$ limit due to the fact that the determinant is
the product over an infinite number of eigenvalues with magnitude
greater than one. Consistently with Eq.~(\ref{eq:66a+++}),
$Det^R[\hat{O}]$ has to be normalized over $Det[\hat{h}]$
($\hat{h}\equiv \, -\partial^2_{\tau} + \omega^2$) which, in the
case of PBC, is: ${Det[\hat{h}]}=\,-4\sinh^2(\omega L/2)$. The
normalization cancels the exponential divergence and makes the
ratio finite.

Then, observing that for $E \rightarrow 0$ ( $T^* \rightarrow 0$):

\begin{eqnarray}
& &W(f_0, f_1)\Bigr|_{\tau_0}\rightarrow {9 \over 8}a^2\omega^2 (1
- p^2)\, \nonumber
\\
& &<f_0 |f_0>\, \rightarrow {6 \over 5}a^2\omega \, \nonumber
\\
& &K(p) \rightarrow \ln(4/\sqrt{1 - p^2})
\label{eq:71}
\end{eqnarray}

from Eq.~(\ref{eq:70}), I finally get the finite ratio

\begin{eqnarray}
{{Det^R[\hat{O}]} \over {Det[\hat{h}]}} \rightarrow  -{{1} \over
{60\omega^2}} \label{eq:71+}
\end{eqnarray}

The dimensionality $[\omega^{-2}]$ correctly accounts for the fact
that one eigenvalue has been extracted from $Det^R[\hat{O}]$. The
computation of Eq.~(\ref{eq:70}) becomes rather time consuming in
the $T^* \rightarrow 0$ limit where elliptic integrals need to be
evaluated with high accuracy in order to get the correct
exponential divergence.

\begin{figure}
\includegraphics[height=7cm,angle=-90]{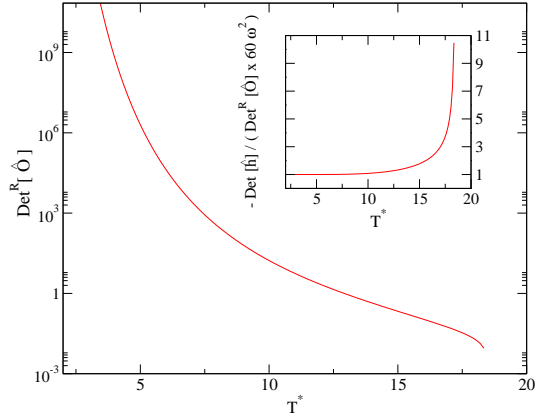}
\caption{\label{fig:5}(Color online) Regularized fluctuation
determinant versus temperature up to $T^*=\,18.339K$. The inset
shows the ratio between harmonic oscillator and regularized
fluctuation determinants. Such ratio is normalized over the
characteristic value $-60\omega^2$ found in the zero temperature
theory. }
\end{figure}

The plot of $Det^R[\hat{O}]$ versus temperature is shown in
Fig.~\ref{fig:5} up to $T^*=\,18.339K < T^*_c$ while the inset
displays the computed inverse ratio normalized over $-60\omega^2$:
the $T^* \rightarrow 0$ limit is in fact an excellent estimate up
to $T^* \sim 10K$ whereas a strong deviation is found at larger
$T^*$ up to $\sim T^*_{c}$. Exactly at $T^*_{c}$, $Det^R[\hat{O}]$
vanishes thus causing the divergence of the inverse ratio. This
divergence may look surprising since the zero mode had been
extracted from the determinant: as it will be unravelled in the
next Subsection, there must be a quantum fluctuation mode which
softens by increasing $T^*$ and ultimately vanishes at the
sphaleron.

\begin{figure}
\includegraphics[height=7.5cm,angle=-90]{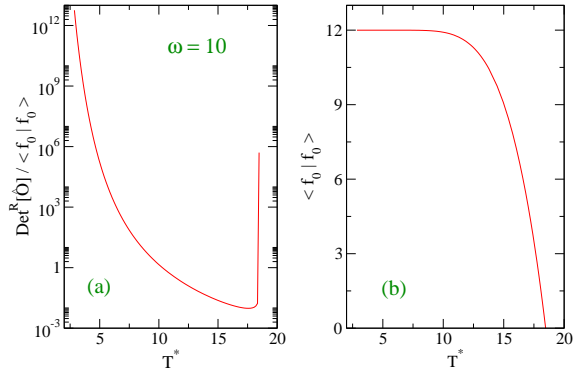}
\caption{\label{fig:6}(Color online) (a) Regularized fluctuation
determinant over norm of the zero eigenmode;  (b) norm of the zero
eigenmode. Both curves are plotted versus temperature including
the crossover point.}
\end{figure}

I want to emphasize the key role played by the bounce velocity
norm in getting $Det^R[\hat{O}] \sim 0$ at $T^* \sim T^*_{c}$:
close to the sphaleron, ${<f_0 | f_0>} \propto p^4$ and $W(f_0,
f_1) \propto p^2$ thus the squared norm drops to zero faster than
any other term in Eq.~(\ref{eq:70}). This feature is made evident
in Fig.~\ref{fig:6} where the plots of ${{Det^R[\hat{O}]}/ {<f_0 |
f_0>}}$ and the squared norm are given separately against
temperature up to $T^*_c$.

Eventually, note that the determinants ratio is negative and this
sign has physical meaning as it is due precisely to the negative
ground state of the fluctuation spectrum. As $Z_1/Z_h$ contributes
to the partition function by the square root of the fluctuation
determinants (inverse) ratio it follows that such contribution is
purely imaginary. To weigh the effect of the negative eigenvalue
$\varepsilon_{-1}$, its analytical expression has to be determined
at finite $T^*$. This is done in the next Subsection.

\subsection*{B. Lam\`{e} Equation}

Let's take Eq.~(\ref{eq:66++}) with the second derivative of the
potential given by

\begin{eqnarray}
{{V''(x_{cl}(\tau))}}=\,{M\omega^2} \Bigl(1 - {2 \over a}
x_{cl}(\tau) \Bigr)  \label{eq:72}
\end{eqnarray}

Then, using Eq.~(\ref{eq:62}) and working out the algebra, I get
the stability equation which governs the fluctuation spectrum

\begin{eqnarray}
& &{{d^2} \over {d\varpi^2}} \eta_n(\tau) =\,\bigl[12 p^2
sn^2(\varpi,p) + \mathcal{A}_n \bigr] \eta_n(\tau) \, \nonumber
\\
& &\mathcal{A}_n=\,{{4(1 - 3\chi_1)} \over {\chi_1 - \chi_3}} -
{{4\varepsilon_n} \over {\omega^2(\chi_1 - \chi_3)}} \, \nonumber
\\
& & 12 \equiv l(l + 1)  \label{eq:73}
\end{eqnarray}

This is the Lam\`{e} equation in the Jacobian form for the case
$l=\,3$ \cite{wang}. For a given $l$ and $p$,
Eq.~(\ref{eq:73}) yields periodic solutions (which can be expanded
in infinite series) for an infinite sequence of characteristic
$\mathcal{A}_n$ values. The continuum of the fluctuation spectrum
stems from this sequence. However, being $l$ positive and integer,
the first $2l + 1$ solutions of Eq.~(\ref{eq:73}) are not infinite
series but polynomials in the Jacobi elliptic functions with real
period $2K(p)$ or $4K(p)$. Being the period of the potential
\cite{ward},  $2K(p)$ plays the role of a lattice constant.

Then,  Eq.~(\ref{eq:73}) admits 7 polynomial solutions with
eigenvalues $\mathcal{A}_n, n\in [-l,l]$, from which the
corresponding $\varepsilon_n$ are derived \cite{n4}:

\begin{eqnarray}
& &\varepsilon_{3}=\,\omega^2\Bigl(\alpha_1 + \alpha_2\Bigl[5(1 +
p^2) + 2\sqrt{4p^4 - 7p^2 + 4}\Bigr] \Bigr) \, \nonumber
\\
& &\varepsilon_{-3}=\,\omega^2\Bigl(\alpha_1 + \alpha_2\Bigl[5(1 +
p^2) - 2\sqrt{4p^4 - 7p^2 + 4}\Bigr] \Bigr) \, \nonumber
\\
& & \varepsilon_{2}=\,\omega^2\Bigl(\alpha_1 + \alpha_2\Bigl[5 +
2p^2 + 2\sqrt{p^4 - p^2 + 4}\Bigr] \Bigr)  \, \nonumber
\\
& &\varepsilon_{-2}=\,\omega^2\Bigl(\alpha_1 + \alpha_2\Bigl[5 +
2p^2 - 2\sqrt{p^4 - p^2 + 4}\Bigr] \Bigr)  \, \nonumber
\\
& & \varepsilon_1=\, \omega^2\Bigl(\alpha_1 + \alpha_2\Bigl[2 +
5p^2 + 2\sqrt{4p^4 - p^2 + 1}\Bigr] \Bigr) \nonumber
\\
& & \varepsilon_{-1}=\,  \omega^2\Bigl(\alpha_1 + \alpha_2 \Bigl[2
+ 5p^2 - 2\sqrt{4p^4 - p^2 + 1}\Bigr] \Bigr)\, \nonumber
\\
& &\varepsilon_0=\,0 \, \nonumber
\\
& &\alpha_1\equiv\, 1 - 3\chi_1  \, \nonumber
\\
& &\alpha_2\equiv\, {{\chi_1 - \chi_3} \over 4} \label{eq:75}
\end{eqnarray}

\begin{figure}
\includegraphics[height=7.5cm,angle=-90]{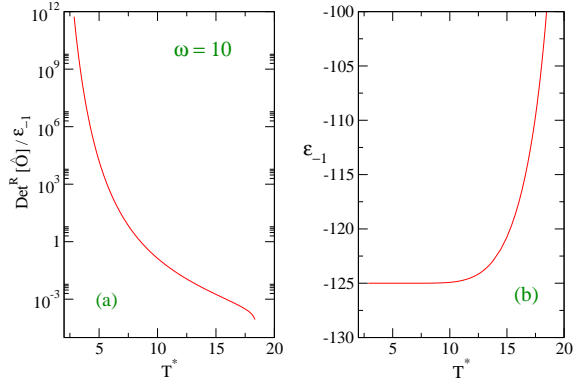}
\caption{\label{fig:7}(Color online) (a) Absolute value of {\it
regularized fluctuation determinant over ground state eigenvalue}
ratio (units $meV^{-4}$) up to $T^*=\,18.339K$; (b) Ground state
fluctuation eigenvalue (units $meV^{2}$) up to $T^*_c$. }
\end{figure}

The physically relevant feature is that the $\varepsilon_n$
depend, through $p$, on the system size $L$ (or on $T^*$). However
not all the $\varepsilon_n$ are good fluctuation eigenvalues. In
fact the first four values have to be discarded as their
eigenfunctions do not fulfill the PBC required for the fluctuation
components: $\eta_n(\varpi_1)=\,\eta_n(\varpi_1 \mp 2K(p))$. Thus,
three good eigenvalues in polynomial form are found: {\it i)}
$\varepsilon_0=\,0$ is the zero mode eigenvalue which is
consistently found also through the stability equation. {\it ii)}
$\varepsilon_1$ lies in the continuum and vanishes at $p=\,0$:
this is the soft mode responsible for the sharp drop observed in
$Det^R[\hat{O}]$ as the sphaleron is approached. At $T_{c}^*$,
$\varepsilon_1$ and $\varepsilon_0$ merge consistently with the
double degeneracy of the corresponding eigenmodes above the
crossover. {\it iii)} $\varepsilon_{-1}$ is the negative
eigenvalue responsible for the decay of the metastable state. In
Fig.~\ref{fig:7}, the ratio $Det^R[\hat{O}]/\varepsilon_{-1}$ and
$\varepsilon_{-1}$ are plotted separately to emphasize the role of
the latter. The dependence on $T^*$ is obtained as usual by means
of Eq.~(\ref{eq:64a}). While the overall shape of $Det^R[\hat{O}]$
is not essentially modified by extracting $\varepsilon_{-1}$
(compare with Fig.~\ref{fig:5}), the interesting feature is
represented by the softening of $\varepsilon_{-1}$ with respect to
the value $\varepsilon_{-1}=\, -5\omega^2/4$ found at $T^*=\,0$.
The substantial reduction (in absolute value) starts up at $T^*
\sim 10K$, that is in the same range at which the classical
properties deviate from the predictions of the zero temperature
theory. Finally, at the sphaleron, from Eq.~(\ref{eq:75}) I get
$\varepsilon_{-1}=\,-\omega^2$ consistently with the observation
made at the end of the previous Section. This completes the
analysis of the quantum fluctuation spectrum.

\section*{4. Decay Rate}

The decay rate in Eq.~(\ref{eq:100}) for the finite temperature theory
can be now computed as the coefficients $A$ and $B$, identified
with $\sqrt{\bigl|Det[\hat{h}]/ Det[\hat{O}]\bigr|}/L$ and
$A[x_{cl}]$ respectively, have been analysed in detail. The
general expression for $\Gamma(T^*)$ is:

\begin{eqnarray}
& & \Gamma(T^*)=\,\hbar  \sqrt{{{M}\over {2\pi\hbar}}}\bar{N}^{-1}
\sqrt{{Det[\hat{h}]} \over {Det^{RR}
[\hat{O}]{\bigl|\varepsilon_{-1}\bigr| }}}
\exp\biggl[-{{A[x_{cl}]} \over {\hbar}} \biggr]  \, \nonumber
\\
\label{eq:76}
\end{eqnarray}

with $Det^{RR} [\hat{O}]\equiv \,Det^R[\hat{O}]/\varepsilon_{-1}$.
Essentially, the two bound state eigenvalues $\varepsilon_{0}$ and
$\varepsilon_{-1}$ have been extracted from $Det[\hat{O}]$ and
treated separately.

It is now clear that, in order to calculate the decay rate, there
would be no need to compute $\bar{N}^{-1}$ due to the zero mode
\cite{tarlie} as it cancels out with the same term embedded in
${Det^R[\hat{O}]}$ and discussed above. Thus, it is the piece of
determinant in Fig.~\ref{fig:6}(a) which enters the computation of
$\Gamma$.

\begin{figure}
\includegraphics[height=7cm,angle=-90]{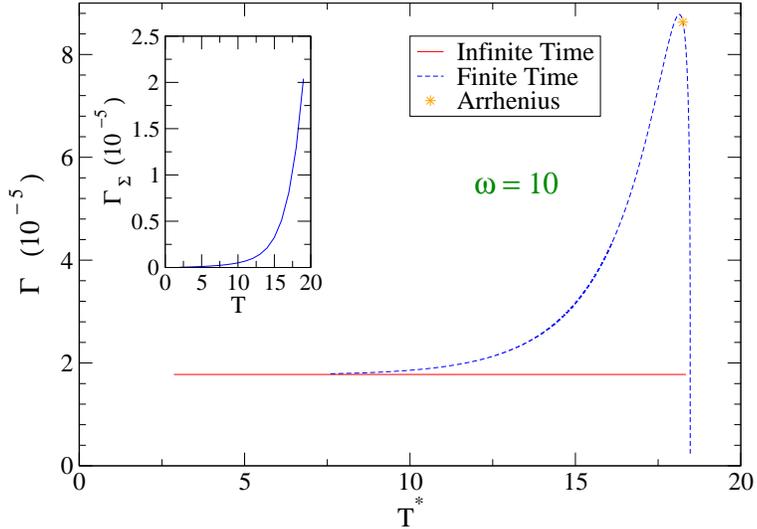}
\caption{\label{fig:8}(Color online) Decay rate (in $meV$) for a
particle in a metastable cubic potential versus temperature up to
$T^*_c$. The result of the standard instantonic  theory is shown for
comparison. The symbol marks the point in which the quantum decay
rate merges with the classical Arrhenius factor. The quantum statistical decay rate $\Gamma_\Sigma$ is
plotted in the inset.}
\end{figure}

Eq.~(\ref{eq:76}) is plotted in Fig.~\ref{fig:8} against
temperature up to $T^*_c=\,18.469K$ together with the constant
decay rate of the {\it zero temperature} theory.  There is a
substantial increase in $\Gamma$ above $T^* \sim 10K$ where the
combined effects of quantum fluctuations and classical action
softening become evident. Approaching the crossover (at
$T_P^*=\,18.138K$), $\Gamma$ reaches a peak value which is larger
by a factor $5$ with respect to the value of the zero $T$
theory although the linewidth remains much smaller than the
oscillator energy, $\Gamma(T_P^*)/\hbar\omega \sim 10^{-5}$. Certainly, such ratio strongly depends on the particle mass through the exponential Gamow factor as ${A[x_{cl}]} \propto M$.
Beyond $T_P^*$ and {\it  before} $T^*_c$, the quantum decay rate
smoothly merges with the classical Arrhenius factor as
${{A[x_{cl}]}/ {\hbar}} \rightarrow V(a)/K_BT^*_c$. In principle
this should occur exactly at $T^*_c$ once the nonuniform bounce
leaves room to the constant solution $x_{cl}(\tau)=\,a$. But close
to $T^*_c$ the decay rate drops to zero as $\Gamma(T^*) \propto
(T^*_c - T^*)^{1/2}$ signalling that the quantum tunneling effects
is ending. The power law dependence is driven by
$\sqrt{N^{-2}/Det^R[\hat{O}]} \propto p$. For this reason the
Arrhenius behavior in the decay rate, which is a continuous function of
$T^*$, takes over slightly before $T^*_c$ and precisely at $T^*_A
=\,18.25K$ corresponding to the symbol in Fig.~\ref{fig:8}.

The inset shows the plot of the quantum statistical decay rate $\Gamma_\Sigma$  \cite{weisshaef} which I have computed, on the base of the WKB turning point formula \cite{affl}, by summing over the particle energies in the cubic potential case. In the quantum regime, the calculation converges by taking about 1000 energy values (below the sphaleron $V(a)$) in the Boltzmann average.  In the activated regime (with energies above the sphaleron), $\Gamma_\Sigma$ continuously grows with exponential behavior while, the matching between the quantum and activated $\Gamma_\Sigma$ plots occurs at around $T_c^*$, with states of energy $E \sim V(a)$ yielding a significant contribution to $\Gamma_\Sigma$ in the classical regime.

The deviation with respect to the $T=\,0$ result found in the
quantum decay rate $\Gamma(T^*)$ (and culminating in the peak value) is
interesting also in view of a comparison with classically
activated systems in which spatio-temporal noise induces
transitions between locally stable states of a nonlinear
potential. The changes in radius and the stability conditions of
metastable metallic nanowires are an example of current interest
\cite{stein1,yanson,stafford}. In these systems, the finite size
shapes the activation behavior and a power-law divergence in the
escape rate with critical exponent $1/2$ is predicted once a
critical lenghtscale is approached.

\begin{figure}
\includegraphics[height=7cm,angle=-90]{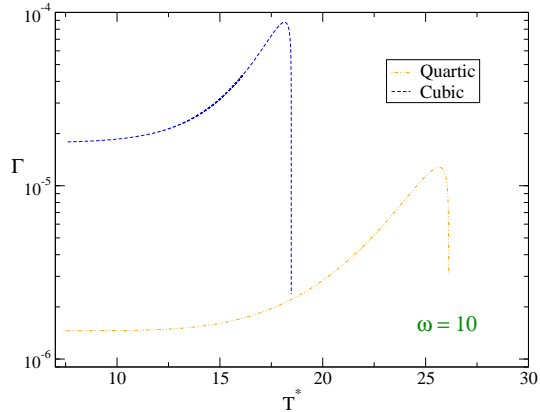}
\caption{\label{fig:9}(Color online) Decay rates of metastable
cubic and quartic potentials versus temperature up to $T^*_c$. }
\end{figure}

In the decay rate of Eq.~(\ref{eq:76}) the small parameter is
$\hbar$ which, unlike the noise in classical systems, cannot be
varied as a function of $T$ (or $L$). Accordingly the quantum
tunneling decay rate is always small. Similar conclusions
regarding the transition can be drawn by carrying out the study of
$\Gamma$ for the metastable quartic potential

\begin{eqnarray}
V(x)=\, {{M\omega^2} \over 2}x^2 - {{\delta} \over 4}x^4
\label{eq:0}
\end{eqnarray}

according to the same semiclassical method described so far. The
parameters $M$, $\omega$ and $a$ have been taken as in the cubic
case while $\delta$, in units $eV \AA^{-4}$, is given by
$\delta=\, {M\omega^2/ a^2}$. There are in fact some fundamental
differences which may be summarized as follows: {\it 1)} for the
quartic potential, the bounce solution is linear in the Jacobi
elliptic $dn$-amplitude and the crossover temperature $T_{c}^*$ is
larger by a factor $\sqrt 2$ with respect to the value given in
Eq.~(\ref{eq:64b}). {\it 2)} The classical action is slightly
larger than that computed in Fig.~\ref{fig:4} and, in the
$E\rightarrow 0$ limit, $A[x_{cl}]\rightarrow \, {{4 M^2 \omega^3}
/ {(3\hbar\delta)}}$. {\it 3)} The {\it harmonic oscillator over
regularized fluctuations determinants} ratio is smaller and, in
the $T^* \rightarrow 0$ limit: ${{Det[\hat{h}]} /
{Det^R[\hat{O}]}} \rightarrow  - {12\omega^2}$. {\it 4)} The
Lam\`{e} equation, which has to be solved for the case $l=\,2$,
has the negative eigenvalue scaling from $\varepsilon_{-1}=\,
-3\omega^2$ at $T^*=\,0$ to $\varepsilon_{-1}=\, -2\omega^2$ at
$T^*=\,T_{c}^*$. All these features work together to decrease the
decay rate of the metastable quartic potential with respect to the
cubic one.   The quantitative results are reported on in
Fig.~\ref{fig:9} by plotting $\Gamma$ for the two potentials. It
appears that, at a given $T^*$ within the quantum regime, the life
of a particle set in a metastable quartic potential is longer (by
an order of magnitude) than in a cubic potential with the same
structural parameters. Close to $T^*_c$, the decay rate of the
quartic potential shows the same behavior found in the cubic case
as a consequence of the softening of the polynomial fluctuation
eigenmode with lowest lying positive eigenvalue.

\section*{5. Conclusion}

The semiclassical path integral method has been applied to compute
the decay rate of a metastable cubic potential at finite
temperature. I have considered a non dissipative system where the
inverse temperature, introduced as the imaginary time of the
Euclidean path integral, represents a measure of the system {\it
size}. Solving the Euler-Lagrange equation for the cubic potential
I have determined the classical background, the bounce, in terms
of Jacobi elliptic functions whose energy dependent modulus keeps
track of the classical mechanics. The periodicity property of the
elliptic solution sets the fundamental relation between bounce
oscillation period and classical energy. As the former varies
monotonically with the latter, the transition from the quantum to
the activated regime is in fact a temperature driven smooth
crossover occuring at $T^*_{c} \propto \omega$. Then the quantum
regime may cover a broader temperature range in systems with
characteristic frequencies larger than the value taken here. In
the latter cases however the quadratic approximation for the
quantum fluctuations may become questionable.

In general, the character of this transition in tunneling systems
may depend on the form of the potential. For the metastable
potential the classical background represented by the bounce has
zero topological charge, fulfills periodic boundary conditions and
the amplitude of its motion progressively shrinks by increasing
$T^*$. At $T^*_{c}$, the bounce degenerates into a point, the
constant solution which determines the classical Arrhenius factor.
Then the nonuniform bounce samples a $T^*$
dependent portion of the (reversed) potential valley and
continuosly adapts its shape to join the turning points. That's
why the crossover is smooth.

As in the semiclassical method classical and quantum degrees of
freedom are intertwinned the quantum fluctuation spectrum can be
evaluated, by the theory of the functional determinants, in terms
of the classical bounce. More precisely, in terms of its time and
modulus derivatives. I have obtained a compact expression for the
fluctuations contribution to the path integral and computed the
renormalization of the fluctuation spectrum due to temperature
effects inside the quantum regime. The overall softening is
remarkable: it shows up at about $T^*_{c}/2$ whereas at lower
temperatures the predictions of the zero $T$ (infinite time)
theory are accurate. The polynomial eigenmodes of the spectrum
have been analytically determined by solving the stability
equation associated to the Sturm-Liouville differential operator,
the Lam\`{e} equation. The focus on the ground state is justified
by the fact that its eigenvalue is negative hence, it is
responsible for the decay rate $\Gamma$ of the metastable
potential. The negative eigenvalue softens considerably by
approaching the sphaleron where, in turn, it determines the
oscillation period of the classical bounce. Also the positive
polynomial eigenvalue plays a key role as it even vanishes at the
sphaleron and ultimately induces the rapid increase in the inverse
determinants ratio. The explicitly derived equations governing the
temperature dependence of the fluctuations may be adapted to
specific situations by tuning the potential parameters. For the
specific choice of parameters here presented, $M=\,10^4m_e$ and
$\hbar\omega=\,10meV$, the computed $\Gamma$ starts to increase
over the zero $T$ decay rate at about $T^*_{c}/2$ and, close to
$T^*_{c}$, the peak value is larger by a factor $5$, a sizeable
effect. At the crossover the quantum decay process ends and the
Arrhenius classical behavior sets in. Finally, a comparison
between cubic and quartic metastable potentials having the same
structural parameters shows that the latter is {\it less
metastable} in the sense that the lifetime is found to be larger
by a factor ten at any temperature in the quantum regime.


\begin{references}
\bibitem{langer}
J.S.Langer, Ann. Phys. {\bf 41}, 108 (1967).
\bibitem{volo}
M.B.Voloshin, I.Y.Kobzarev, L.B.Okun, Sov. J. Nucl. Phys. {\bf
20}, 644 (1975).
\bibitem{bender}
T.Banks, C.Bender, T.T.Wu, Phys.\ Rev.\ D {\bf 8}, 3346 (1973).
\bibitem{cole}
S.Coleman, Phys.\ Rev.\ D {\bf 15},
2929 (1977); C.G.Callan, S.Coleman, Phys.\ Rev.\ D {\bf 16}, 1762
(1977).
\bibitem{landau}
L.D.Landau, E.M.Lifshitz, {\it Quantum Mechanics} $3^{rd}$ edn.
(Butterworth-Heinemann, Oxford, 1977).
\bibitem{feyn}
R.P.Feynman, Rev.\ Mod.\ Phys. {\bf 20}, 367 (1948).
\bibitem{schul}
L.S.Schulman  {\it Techniques and Applications of Path
Integration} (Wiley\&Sons, New York, 1981).
\bibitem{gelfand}
I.M.Gelfand, A.M.Yaglom, J.Math.Phys. {\bf 1}, 48 (1960).
\bibitem{zoli}
M.Zoli,  Phys. Rev. B, {\bf 72}, 214302 (2005).
\bibitem{rajara}
R.Rajaraman, {\it Solitons and Instantons} (North Holland,
Amsterdam, 1982).
\bibitem{parisi}
E.Br\'{e}zin, G.Parisi, J.Zinn-Justin, Phys.\ Rev.\ D {\bf 16},
408 (1977).
\bibitem{affl}
I.Affleck, Phys.\ Rev.\ Lett. {\bf 46}, 388 (1981).
\bibitem{weisshaef}
U.Weiss, W.Haeffner, Phys.\ Rev.\ D {\bf 27}, 2916 (1983).
\bibitem{volo1}
M.B.Voloshin, Phys.\ Lett.\ B {\bf 599}, 129 (2004).
\bibitem{kleinert}
H.Kleinert, {\it Path Integrals in
Quantum Mechanics, Statistics, Polymer Physycs and Financial
Markets} (World Scientific Publishing, Singapore, 2004).
\bibitem{braun}
H.-B.Braun,  Phys.\ Rev.\ Lett. {\bf 71}, 3557 (1993).
\bibitem{stein1}
J.B\"{u}rki, C.A.Stafford, D.L.Stein,  Phys.\ Rev.\ Lett. {\bf
95}, 090601 (2005).
\bibitem{faris}
W.G.Faris, G.Jona-Lasinio, J.\ Phys.\ A {\bf 15}, 3025 (1982).
\bibitem{n1}
The concept of temperature needs a clarification. In theories of
phase transitions with symmetry breaking the effective potential
is a function of the temperature. By cooling the system, a minimum
(corresponding to a non-zero Higgs field $\varphi$) appears in the
potential. Thus the metastable state placed at $\varphi=\,0$
decays via quantum tunneling to the $\varphi \neq\,0$
energetically favourable state. Here I'm taking a specific
potential which does not change its shape with $T$: it has a
metastable state at any $T$ within the considered quantum regime.
Instead, it is the particle moving in such potential which feels
the temperature effects. The bounce plus its quantum
fluctuations depend on $T$ which is here an inverse {\it time}.
\bibitem{wang}
Z.X.Wang, D.R.Guo, {\it Special Functions} (World
Scientific, Singapore, 1989).
\bibitem{hanggi}
P.H\"{a}nggi, P.Talkner, M.Borkovec, Rev.\ Mod.\ Phys. {\bf 62},
251 (1990).
\bibitem{antunes}
N.Antunes, F.C.Lombardo, D.Monteoliva, P.I.Villar, Phys.\ Rev.\ E
{\bf 73}, 066105 (2006).
\bibitem{rise}
P.S.Riseborough, P.H\"{a}nggi, E.Freidkin, Phys.\ Rev.\ A {\bf
32}, 489 (1985).
\bibitem{stein2}
D.L.Stein, Brazilian J. Phys.  {\bf 35}, 242 (2005).
\bibitem{manton}
N.S.Manton, Phys.\ Rev.\ D {\bf 28}, 2019 (1983).
\bibitem{grad}
I.S.Gradshteyn, I.M.Ryzhik, {\it Tables of Integrals, Series and
Products} (Academic Press, New York, 1965).
\bibitem{gold}
 V.I.Goldanskii, Sov.Phys.Dokl. {\bf 4}, 74  (1959).
\bibitem{larkin}
A.I.Larkin,  Y.N.Ovchinnikov, Sov.Phys.JETP {\bf 37}, 382 (1983);
ibid. {\bf 59}, 420 (1984).
\bibitem{grabwei}
H.Grabert, U.Weiss, Phys.\ Rev.\ Lett. {\bf 53}, 1787 (1984).
\bibitem{blatter}
D.A.Gorokhov, G.Blatter, Phys.\ Rev.\ B {\bf 56}, 3130 (1997).
\bibitem{kuznet}
A.N.Kuznetsov, P.G.Tinyakov, Phys. Lett. B {\bf 406}, 76 (1997).
\bibitem{liang}
J.-Q.Liang, H.J.W.M\"{u}ller-Kirsten, D.K.Park, F.Zimmerschied,
Phys.\ Rev.\ Lett. {\bf 81}, 216 (1998).
\bibitem{chudno}
E.M.Chudnovski, Phys.\ Rev.\ A {\bf 46}, 8011 (1992).
\bibitem{n3}
This can be directly deduced from Eq.~(\ref{eq:64}) as
$K(p=0)=\,\pi/2$.
\bibitem{park}
S.Y.Lee, H.Kim, D.K.Park, J.K.Kim, Phys.\ Rev.\ B {\bf 60}, 10086
(1999).
\bibitem{forman}
R.Forman, Invent.Math. {\bf 88}, 447 (1987); ibid. Commun.\ Math.\
Phys. {\bf 147}, 485 (1992).
\bibitem{kirsten1}
K.Kirsten, A.J.McKane, J.Phys A:Math. Gen. {\bf 37}, 4649 (2004).
\bibitem{ward}
R.S.Ward, J.Phys.A: Math.Gen. {\bf 20}, 2679 (1987).
\bibitem{n4}
The way to order eigenvalues and eigenstates in terms of $n$ is
purely conventional. Either $n \in [-l,l]$ or $n \in [0,2l + 1]$
can be used. The former is assumed here consistently with having
labelled the unstable eigenvalue by $n=\,-1$.
\bibitem{tarlie}
A.J.McKane, M.B.Tarlie, J.Math.Phys. {\bf 28}, 6931 (1995).
\bibitem{yanson}
A.I.Yanson, I.K.Yanson, J.M. van Ruitenbeck, Nature {\bf 400}, 144
(1999).
\bibitem{stafford}
C.-H.Zhang, F.Kassubek, C.A.Stafford,  Phys.\ Rev.\ B {\bf 68},
165414 (2003).



\end{references}
\end{document}